\documentclass{modelica}
\usepackage[draft=true]{minted}

\hypersetup{%
  pdftitle  = {Composing Modeling and Simulation with \\Machine Learning in Julia},
  pdfauthor = {Chris Rackauckas, Ranjan Anantharaman, Alan Edelman, Shashi Gowda, Maja Gwozdz, Anand Jain, Chris Laughman, Yingbo Ma, Francesco Martinuzzi, Avik Pal, Utkarsh Rajput, Elliot Saba, Viral B. Shah},
 pdfsubject = {14th International Modelica Conference 2021},
 pdfkeywords = {Modelica, conference, LaTeX, template},
 colorlinks,
 linkcolor=black,
 urlcolor=black,
 citecolor=black,
 pdfpagelayout = SinglePage,
 pdfcreator = pdflatex,
 pdfproducer = pdflatex}

\begin{document}
\thispagestyle{empty}

\title{Composing Modeling and Simulation with \\Machine Learning in Julia}
\author[1,2]{Chris Rackauckas}
\author[2]{Ranjan Anantharaman}
\author[2]{Alan Edelman}
\author[2]{Shashi Gowda}
\author[1]{Maja Gwozdz}
\author[1]{Anand Jain}
\author[3]{Chris Laughman}
\author[1]{Yingbo Ma}
\author[1]{Francesco Martinuzzi}
\author[1]{Avik Pal}
\author[1]{Utkarsh Rajput}
\author[1]{Elliot Saba}
\author[1]{Viral B. Shah}

\affil[1]{Julia Computing Inc., USA}
\affil[2]{Massachusetts Institute of Technology, USA}
\affil[3]{Mitsubishi Electric Research Lab, USA}

% \title{\textbf{Int. Modelica Conf. 2021 Paper Title}}
% \author{{\large
% Author Name$^1$ \quad Author Name$^1$ \quad Author Name$^2$\vspace{2mm}\\
%   {}$^1$Department, University, Country, \textsf{\{name1,name2\}@university.org}\\
%   {}$^2$Company, Country, \textsf{name3@company}}

\maketitle\thispagestyle{empty} %% <-- you need this for the first page
\abstract{%
In this paper we introduce JuliaSim, a high-performance programming environment designed to blend traditional modeling and simulation with machine learning. JuliaSim can build accelerated surrogates from component-based models, such as those conforming to the FMI standard, using continuous-time echo state networks (CTESN). The foundation of this environment, ModelingToolkit.jl, is an acausal modeling language which can compose the trained surrogates as components within its staged compilation process. As a complementary factor we present the JuliaSim model library, a standard library with differential-algebraic equations and pre-trained surrogates, which can be composed using the modeling system for design, optimization, and control. We demonstrate the effectiveness of the surrogate-accelerated modeling and simulation approach on HVAC dynamics by showing that the CTESN surrogates accurately capture the dynamics of a HVAC cycle at less than 4\% error while accelerating its simulation by 340x. We illustrate the use of surrogate acceleration in the design process via global optimization of simulation parameters using the embedded surrogate, yielding a speedup of two orders of magnitude to find the optimum. We showcase the surrogate deployed in a co-simulation loop, as a drop-in replacement for one of the coupled FMUs, allowing engineers to effectively explore the design space of a coupled system. Together this demonstrates a workflow for automating the integration of machine learning techniques into traditional modeling and simulation processes.
}


\section{Introduction}

% Survey surrogates for modelica environments and simulation environments

With the dramatic success of artificial intelligence and machine learning (AI/ML) throughout many disciplines, one major question is how AI/ML will change the field of modeling and simulation. Modern modeling and simulation involves the time integration of detailed multi-physics component models, programmatically generated by domain-specific simulation software. Their large computational expense makes design, optimization and control of these systems prohibitively expensive \cite{benner2015survey}. Thus one of the major proposed avenues for AI/ML in the space of modeling and simulation is in the generation of reduced models and data-driven surrogates, that is, sufficiently accurate approximations with majorly reduced computational burden \cite{willard2020integrating, ratnaswamy2019physics, zhang2020hydrological, kim2020fast, hu2020neural}. While the research has shown many cross-domain successes, the average modeler does not employ surrogates in most projects for a number of reasons: the surrogatization process is not robust enough to be used blindly, it can be difficult to ascertain whether the surrogate approximation is sufficiently accurate to trust the results, and it is not automated in modeling languages. This begs the question -- how does one develop a modeling environment that seamlessly integrates traditional and machine learning approaches in order to merge this newfound speed with the robustness of stabilized integration techniques?

The difficulty of addressing these questions comes down to the intricate domain-specific algorithms which have been developed over the previous decades. Many scientists and engineers practice modeling and simulation using acausal modeling languages, which require sophisticated symbolic algorithms in order to give a stable result. Algorithms, such as alias elimination \cite{otter2017transformation} and the Pantelides algorithm for index reduction \cite{Pantelides:1988}, drive the backend of current Modelica compilers like Dymola \cite{bruck2002dymola} and OpenModelica \cite{fritzson2005openmodelica} and allow for large-scale differential-algebraic equation (DAE) models to be effectively solved. Notably, these compiler pipelines encode exact symbolic transformations. One can think of generalizing this process by allowing approximate symbolic transformations, which can thus include model reduction and machine learning techniques. As this process now allows for inexact transformation, the modeling language would need to allow users to interact with the compiler. Moreover, it would have to allow users to swap in and out approximations, selectively accelerate specific submodels, and finally make it easy to check the results against the non-approximated model.

To address these issues, we introduce JuliaSim --- a modeling and simulation environment, which merges elements of acausal modeling frameworks like Modelica with machine learning elements. The core of the environment is the open source ModelingToolkit.jl \cite{ma2021modelingtoolkit}, an acausal modeling framework with an interactive compilation mechanism for including exact and inexact transformations. To incorporate machine learning, we describe the continuous-time echo state network (CTESN) architecture as an approximation transformation of time series data to a DAE component. Notably, the CTESN architecture allows for an implicit training to handle the stiff equations common in engineering simulations. To demonstrate the utility of this architecture, we showcase the CTESN as a methodology for translating a Room Air Conditioner model from a Functional Mock-up Unit (FMU) binary to an accelerated ModelingToolkit.jl model with 4\% error over the operating parameter range, accelerating it by 340x. We then show how the accelerated model can be used to speed up global parameter optimization by over two orders of magnitude. As a component within an acausal modeling framework, we demonstrate its ability to be composed with other models, here specifically in the context of the FMI co-simulation environment.

\section{Overview of JuliaSim}

\begin{figure*}[t]
    \centering
    \includegraphics[width=0.6\textwidth]{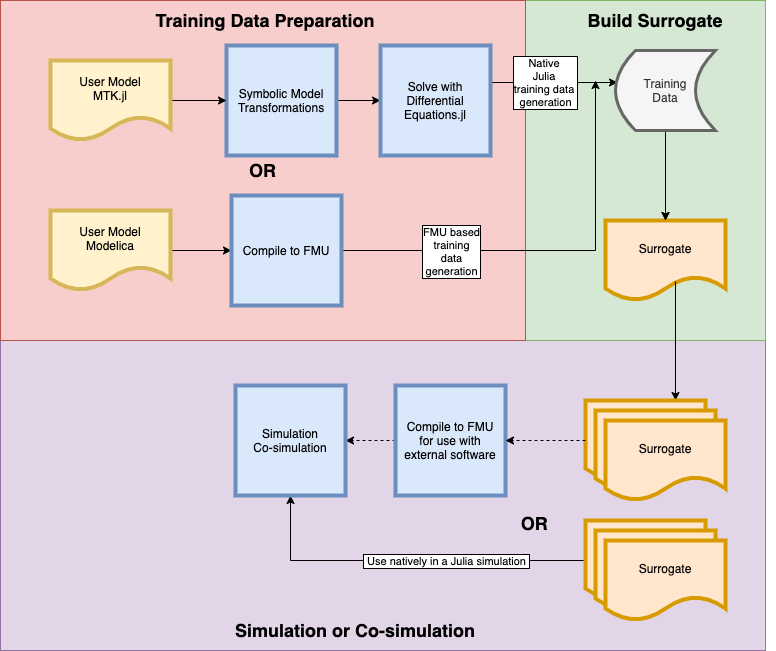}
    \caption{Compiler passes in the JuliaSim Modeling and Simulation system. Ordinarily, most systems simulate equation-based models, described in the ``Training Data Preparation'' and the ``Simulation or Co-simulation'' phases. We provide an additional set of steps in our compiler to compute surrogates of models. Blue boxes represent code transformations, yellow represents user source code, gray represents data sources, and gold represents surrogate models. The dotted line indicates a feature that is currently work in progress.}
    \label{fig:juliasim}
\end{figure*}

The flow of the architecture (Figure~\ref{fig:juliasim}) is described as follows. We start by describing the open ModelingToolkit.jl acausal modeling language as a language with composable transformation passes to include exact and approximate symbolic transformations. To incorporate machine learning into this acausal modeling environment, we describe the CTESN, which is a learnable DAE structure that can be trained on highly stiff time series to build a representation of a component. To expand the utility of components, we outline the interaction with the FMI standard to allow for connecting and composing models. Finally, we present the JuliaSim model library, which is a collection of acausal components that includes pre-trained surrogates of models so that users can utilize the acceleration without having to pay for the cost of training locally.

\subsection{Interactive Acausal Modeling with ModelingToolkit.jl}

ModelingToolkit.jl \cite{ma2021modelingtoolkit} (MTK) is a framework for equation-based acausal modeling written in the Julia programming language \cite{bezanson2017julia}, which generates large systems of DAEs from symbolic models. Similarly to Modelica, it allows for building models hierarchically in a component-based fashion. For example, defining a component in MTK is to define a function which generates an ODESystem:

\begin{minted}[fontsize=\small]{julia}
function Capacitor(;name, C = 1.0)
    val = C
    @named p = Pin(); @named n = Pin()
    @variables v(t); @parameters C
    D = Differential(t)
    eqs = [v ~ p.v - n.v
           0 ~ p.i + n.i
           D(v) ~ p.i / C]
    ODESystem(eqs, t, [v], [C],
        systems=[p, n], 
        defaults=Dict(C => val),
        name=name)
end
\end{minted}

Systems can then be composed by declaring subsystems and defining the connections between them. For instance, the classic RC circuit can be built from standard electrical components as:

\begin{minted}[fontsize=\small]{julia}
@named resistor = Resistor(R=100)
@named capacitor = Capacitor(C=0.001)
@named source = ConstantVoltage(V=10)
@named ground = Ground()
@named rc_model = ODESystem([
    connect(source.p, resistor.p)
    connect(resistor.n, capacitor.p)
    connect(capacitor.n, source.n, 
            ground.g)],
    t, systems=[resistor, capacitor, 
                source, ground])
\end{minted}

The core of MTK's utility is its system of transformations, where a transformation is a function which takes an \verb AbstractSystem  type to another \verb AbstractSystem  type. Given this definition, transformations can be composed and chained. Transformations, such as \texttt{dae\_index\_lowering}, transform a higher-index DAE into an index-1 DAE via the Pantelides algortithm \cite{Pantelides:1988}. Nonlinear tearing and \texttt{alias\_elimination} \cite{otter2017transformation} are other commonly used transformations, which match the workflow of the Dymola Modelica compiler \cite{bruck2002dymola} (and together are given the alias \texttt{structural\_simplify}). However, within this system the user can freely compose transformations with domain- and problem-specific transformations, such as ``exponentiation of a variable to enforce positivity'' or ``extending the system to include the tangent space''. After transformations have been composed, the \verb ODEProblem  constructor compiles the resulting model to a native Julia function for usage with DifferentialEquations.jl \cite{rackauckas2017differentialequations}.

\subsection{Representing Surrogates as DAEs with Continuous-Time Echo State Networks}

In order to compose a trained machine learning model with the components of ModelingToolkit.jl, one needs to represent such a trained model as a set of DAEs. To this end, one can make use of continuous machine learning architectures, such as neural ODEs \cite{chen2018neural} or physics-informed neural networks \cite{raissi2019physics}. However, prior work has demonstrated that such architectures are prone to instabilities when being trained on stiff models~\cite{wang2020understanding}. In order to account for these difficulties, we have recently demonstrated a new architecture, CTESNs, which allows for implicit training in parameter space to stabilize the ill-conditioning present in stiff systems \cite{anantharaman2020accelerating}. For this reason, CTESNs are the default surrogate algorithm of JuliaSim and will be the surrogate algorithm used throughout the rest of the paper.

The CTESN is a continuous-time generalization of echo state networks (ESNs) \cite{lukovsevivcius2012practical}, a reservoir computing framework for learning a nonlinear map by projecting the inputs onto high-dimensional spaces through predefined dynamics of a nonlinear system \cite{lukovsevivcius2009reservoir}. CTESNs are effective at learning the dynamics of systems with widely separated time scales because their design eliminates the requirement of training via local optimization algorithms, like gradient descent, which are differential equation solvers in a stiff parameter space. Instead of using optimization, CTESNs are semi-implicit neural ODEs where the first layer is fixed, which results in an implicit training process.

To develop the CTESN, first a non-stiff dynamical system, called the reservoir, is chosen. This is given by the expression

\begin{gather}
r' = f\left(Ar + W_{hyb} x\left(p^*, t\right)\right) \label{eq:lpctesn_eq1}
\end{gather}

where $A$ is a fixed random sparse $N_R \times N_R$ matrix, $W_{hyb}$ is a fixed random dense $N_R \times N$ matrix, and $x(p^*, t)$ is a solution of the system at a candidate set of parameters from the parameter space, and $f$ is an activation function. 

Projections ($W_{out}$) from the simulated reservoir time series to the truth solution time series are then computed, using the following equation: 

\begin{gather}
    x(t) = g\left(W_{out}r(t)\right) \label{eq:projection}
\end{gather}

where $g$ is an activation function (usually the identity), $r(t)$ represents the solution to the reservoir equation, and $x(t)$ represents the solution to full model. This projection is usually computed via least-squares minimization using the singular value decomposition (SVD), which is robust to ill-conditioning by avoiding gradient-based optimization. A projection is computed for each point in the parameter space, and a map is constructed from the parameter space $P$ to each projection matrix $W_{out}$ (in our examples, we will use a radial basis function to construct this map). Thus our final prediction is the following:

\begin{gather}
    \hat{x}(t) = g(W_{out}(\hat{p})r(t))
\end{gather}

For a given test parameter $\hat{p}$, a matrix $W_{out}(\hat{p})$ is computed, the reservoir equation is simulated, and then the final prediction $\hat{x}$ is a given by the above matrix multiplication.

% Stiff Systems generally have fast transients where gradient computation is highly unstable. Reservoir Computing frameworks like ESNs directly solve for $W_{out}$ using QR Factorization and don't rely on gradient computation. Using a direct linear solve makes them a suitable candidate for surrogatizing systems having gradient pathologies. Additionally, since stiff systems contain rapidly and slowly varying components over multiple timescales, we cannot use uniformly spaced time points. We use CTESN as the default surrogatization algorithm due to its ability to handle both stable slow reacting systems and fast transients in stiff systems~\cite{anantharaman2020accelerating}.

% JuliaSim currently supports two variants of the CTESN framework -- Linear Projection CTESN (LPCTESN) and Non-Linear Projection CTESN (NPCTESN). We define LPCTESN with $N_R$ dimensional reservoir as:
%

%
% and they are set to $tanh$ and $identity$ respectively in our experiments, unless specified otherwise. This formulation allows us to learn the $W_{out}$ matrix by global $L_2$ fitting via stabilized methods like SVD. The prediction $\hat{x}(t)$ is made using an interpolated $W_{out}(\hat{p})$ matrix.

While the formulation above details linear projections from the reservoir time series (Linear Projection CTESN or LPCTESN), nonlinear projections in the form of parametrized functions can also be used to project from the reservoir time series to the reference solution (Nonlinear Projection CTESN). For this variation, a radial basis function can be applied to model the nonlinear projection $r(t) \mapsto x(t)$ in equation \ref{eq:projection}. The learned polynomial coefficients $\beta_i$ from radial basis functions are used, and a mapping between the model parameter space and coefficients $\beta_i$'s is constructed.
\begin{gather}
     \text{rbf}(\beta_i)(r(t)) \approx x(p_i, t) \quad \forall i \in \{1, \dots, k\}\\
     \text{rbf}(p_i) \approx \beta_i \quad \forall i \in \{1, \dots, k\}
\end{gather}

where $k$ is the total number of parameter samples used for training. Finally, during prediction, first the coefficients are predicted and a radial basis function for the prediction of the time series is constructed:
 \begin{gather}
    \hat{\beta} = \text{rbf}(\hat{p})\\
    \hat{x}(t) = \text{rbf}(\hat{\beta})(r(t))
\end{gather}
Notice that both the LPCTESN and the NPCTESN represent the trained model as a set of DAEs, and thus can be represented as an \verb ODESystem  in MTK, and can be composed similarly to any other DAE model.

%
%npctesn vs lpctesn:
%rober example 
%lpctesn JuliaSim.relerror(truth, pred) = [0.013508631203897947, 0.4166474072369464, 0.015129202054309553] = 0.1484 
%training time 295.264066 seconds
%pred time 1.366712 seconds
%npctesn JuliaSim.relerror(truth, pred) = [0.01989083597755903, 0.017832253382790896, 0.02227791738377206] = 0.0200
%training time 466.259363 seconds
%pred time 0.711241 seconds

A significant advantage of applying NPCTESNs over LPCTESNs is the reduction of reservoir sizes, which creates a cheaper surrogate with respect to memory usage. LPCTESNs often use reservoirs whose dimensions reach an order of 1000. While this reservoir ODE is not-stiff, and is cheap to simulate, this leads to higher memory requirements. Consider the surrogatization of the Robertson equations \cite{robertson1976numerical}, a canonical stiff benchmark problem:

\begin{align} \label{eq:robert}
\dot{y_1} &= -0.04 y_1+10^{4} y_2 \cdot y_3 \\
\dot{y_2} &= 0.04 y_1-10^{4} y_2 \cdot y_3-3 \cdot 10^{7} y_2^{2} \\
\dot{y_3} &= 3 \cdot 10^{7} y_2^{2}
\end{align}

where $y_1$, $y_2$, and $y_3$ are the concentrations of three reagants. This system has widely separated reaction rates ($0.04, 10^{4}, 3\cdot10^{7}$), and is well-known to be very stiff \cite{gobbert1996robertson, robertson1975some, robertson1976numerical}. It is commonly used as an example for evaluating integrators of stiff ODEs \cite{hosea1996analysis}. Finding an accurate surrogate for this system is difficult because it needs to capture both the stable slow-reacting system and the fast transients. This breaks many data-driven surrogate methods, such as PINNs and LSTMs \cite{anantharaman2020accelerating}.

Table \ref{tab:lpnp} shows the result of surrogatization using the LPCTESN and the NPCTESN, while considering the following ranges of design parameters corresponding to the three reaction rates: $(0.036, 0.044)$, $(2.7 \cdot 10^{7}, 3.3 \cdot 10^{7})$ and $(0.9 \cdot 10^{4}, 1.1 \cdot 10^{4})$. We observe three orders of magnitude smaller reservoir equation size, resulting in a computationally cheaper surrogate model.

% Even thought the NPCTESN results to be less space demanding when compared to the LPCTESN, the latter shows a faster training process, since it relies on matrix multiplications instead of function evaluations. In the example analyzed above we LPCTESN took 295.2640 seconds compared to the 466.2593 seconds needed for the NPCTESN.

 \begin{table}[htbp]
   \caption{Comparison between LPCTESN and NPCTESN on surrogatization of the Robertson equations. ``Res'' stands for reservoir.}\label{tab:lpnp}
   \centering
  \begin{tabular}{ccc} \toprule
       \emph{Model} & \emph{Res. ODE size} & \emph{Avg Rel. Err \%} \\
       LPCTESN & 3000 & 0.1484 \\ 
       NPCTESN & 3 & 0.0200 \\
       \bottomrule
   \end{tabular}
\end{table}

\begin{figure*}[htbp]
    \centering
    \includegraphics[scale=0.5]{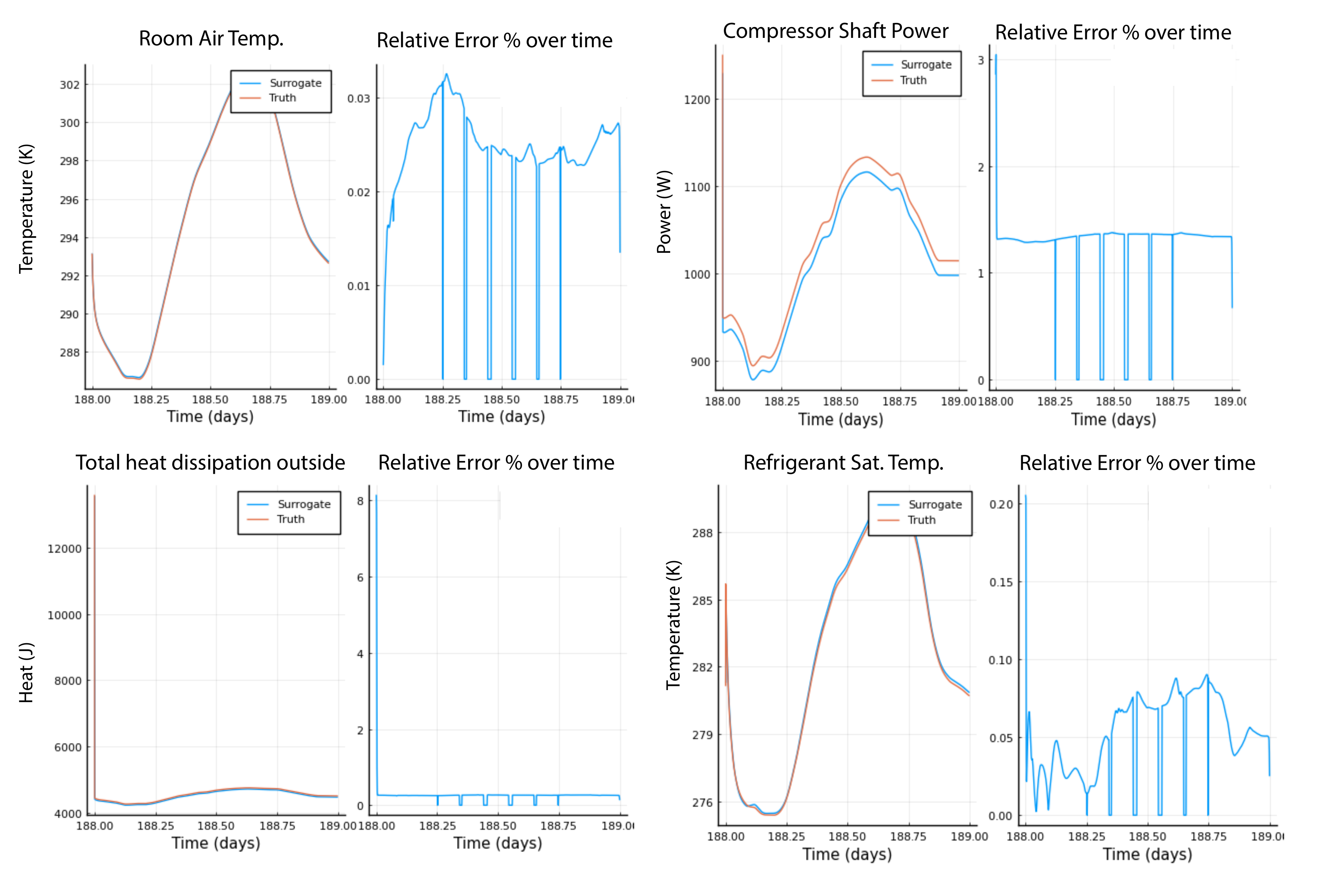}
    \caption{Surrogate prediction of the room temperature of the RAC model in blue, while the ground truth is in red. This is a prediction for points over which the surrogate has not been trained. Relative error is calculated throughout the time span at 1000 uniformly spaced points. The CTESN surrogate was trained on a timespan of an entire day, using data from 100 simulations. The simulation parameters were sampled from a chosen input space using Latin hypercube sampling. The simulation time span goes from 188 days to 189 days at a fixed step size of 5 seconds. Table \ref{tab:inputs} presents the list of and ranges of inputs the surrogate has been trained on. The relative error usually peaks at a point with a discontinuous derivative in time, usually induced by a step or ramp input (which, in this case, is the parametrized compressor speed ramp input.). Another feature of the prediction error above is that it is sometimes stable throughout the time span (such as with the compressor shaft power, top right). This is a feature of how certain outputs vary through the parameter space. Sampling the space with more points or reducing the range of the chosen input space would reduce this error. Table \ref{tab:testtab} shows the maximum relative error computed for many other outputs of interest. Figure \ref{fig:racroom_tests} computes and aggregates maximum errors across a 100 new test points from the space. }
    \label{fig:racroom_validate}
\end{figure*}

\begin{figure*}
    \centering
    \includegraphics[width=0.4\textwidth]{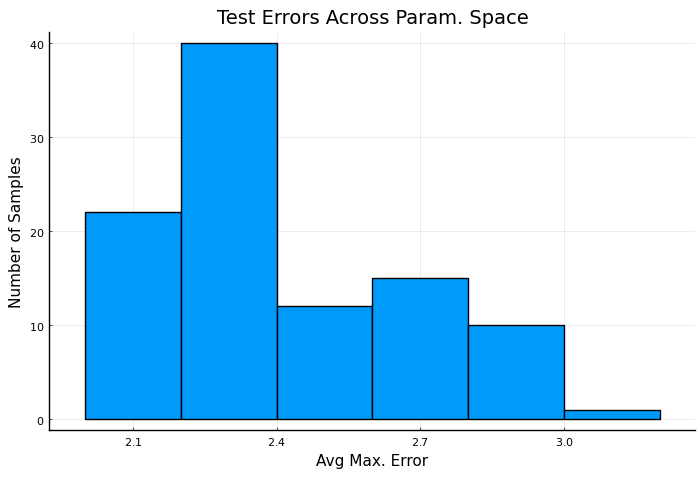}
    \caption{Performance of surrogate when tested on 100 test parameters from the parameter space. The test parameters were chosen via Sobol low discrepancy sampling, and maximum relative error across the time span was calculated for all output quantities. The average maximum error across all output quantities was then plotted as a histogram. Our current test points may not be maximally separated  through the space, but we anticipate similar performance with more test examples and a maximal sampling scheme.}
    \label{fig:racroom_tests}
\end{figure*}
\vspace{-10px}

 \begin{table*}[htbp]
   \caption{Relative errors when the surrogate is tested on parameters it has not been trained on. HEX stands for ``heat exchanger'' and LEV stands for ``linear expansion valve''.}\label{tab:testtab}
   \centering
  \begin{tabular}{lrlr} \toprule
       \emph{Output quantity} & \emph{Max. Rel. Err \%} & \emph{Output quantity} & \emph{Max. Rel. Err \%}\\
       \midrule
      Air temp. in room & {0.033} &
        Rel. humidity in room & 0.872 \\
      Outdoor dry bulb temp. & 0.0001 &
       Outdoor rel. humidity & 0.003 \\
       Compressor inlet pressure & 4.79 &
       Compressor outlet pressure & 3.50 \\
       LEV inlet pressure & 3.48 &
       LEV outlet pressure & 4.84 \\
       LEV refrigerant outlet enthalpy & 1.31 &
       Compressor refrigerant mass flow rate & 4.51 \\
       Evaporator refrigerant saturation temp. & 0.205 &
       Evaporator refrigerant outlet temp. & 0.145 \\
       Total heat dissipation of outdoor HEX & 8.15 &
       Sensible heat load of indoor HEX & 0.892 \\ % incorrect
       Latent heat load of indoor HEX & 3.51 & % incorrect
       Outdoor coil outlet air temperature & 0.432 \\
       Indoor coil outlet air temperature & 0.070 &
       Compressor shaft power & 3.04 \\
       \bottomrule
   \end{tabular}
\end{table*}

 \begin{table}[htbp]
   \caption{Surrogate Operating Parameters. The surrogate is expected to work over this entire range of design parameters.}\label{tab:inputs}
   \centering
  \begin{tabular}{lr} \toprule
       \emph{Input} & \emph{Parameter Range}\\
       \midrule \\
       Compressor Speed (ramp) & Start Time - (900, 1100) s \\ 
        & Start Value - (45, 55) rpm \\
        & Offset - (9, 11) rpm \\
       LEV Position & (252, 300) \\
       Outdoor Unit Fan Speed & (680, 820) rpm \\
       Indoor Unit Fan Speed & (270, 330) rpm \\
       Radiative Heat Gain & (0.0, 0.1) \\ 
       Convective Heat Gain & (0.0, 0.1) \\ 
       Latent Heat Gain & (0.3, 0.4) \\
       \bottomrule
   \end{tabular}
\end{table}

\subsection{Composing with External Models via the FMI Standard}

While these surrogatized CTESNs can be composed with other MTK models, more opportunities can be gained by composing with models from external languages. The Functional Mock-up Interface (FMI) \cite{blochwitz2011functional} is an open-source standard for coupled simulation, adopted and supported by many simulation tools\footnote{https://fmi-standard.org/tools/}, both open source and commercial. Models can be exported as \emph{Functional Mock-up Units} (FMUs), which can then be simulated in a shared environment. Two forms of coupled simulation are standardized. \emph{Model exchange} uses a centralized time-integration algorithm to solve the coupled sets of differential-algebraic equations exported by the individual FMUs. The second approach, \emph{co-simulation}, allows FMUs to export their own simulation routine, and synchronizes them using a master algorithm. Notice that as DAEs, the FMU interface is compatible with ModelingToolkit.jl components and, importantly, trained CTESN models.

JuliaSim can simulate an FMU in parallel at different points in the design space. For each independent simulation, the \texttt{fmpy} package\footnote{https://github.com/CATIA-Systems/FMPy} was used to run the FMU in ModelExchange with CVODE~\cite{cohen1996cvode} or co-simulation with the FMUs exported solver. The resultant time series was then fitted to cubic splines. Integration with state-of-the-art solvers from  DifferentialEquations.jl~\cite{rackauckas2017differentialequations} for simulating ModelExchange FMUs is planned in future releases. 

\subsection{Incorporating Surrogates into the JuliaSim Model Library}

% \subsection{Automated Surrogate Generation}

% \todo[inline]{(Avik) Might be good to get \textbf{Anand's} help for this section on how CellML models are automatically surrogatized}

Reduced order modeling and surrogates in the space of simulation have traditionally targeted PDE problems because of the common reuse of standard PDE models such as Navier-Stokes equations. Since surrogates have a training cost, it is only beneficial to use them if that cost is amortized over many use cases. In equation-based modeling systems, such as Modelica or Simulink, it is common for each modeler to build and simulate a unique model. While at face value this may seem to defeat opportunities for amortizing the cost, the composability of components within these systems is what grants a new opportunity. For example, in Modelica it is common to hierarchically build models from components originating in libraries, such as the Modelica standard library. This means that large components, such as high-fidelity models of air conditioners, specific electrical components, or physiological organelles, could be surrogatized and accelerate enough workflows to overcome the training cost\footnote{We note that an additional argument can be made for pre-trained models in terms of user experience. If a user of a modeling software needs a faster model for real-time control, then having raised the total simulation cost to reduce the real-time user cost would still have a net benefit in terms of the application}. In addition, if the modeler is presented with both the component and its pre-trained surrogate with known accuracy statistics, such a modeler could effectively use the surrogate (e.g., to perform a parameter study) and easily swap back to the high- fidelity version for the final model.

Thus to complement the JuliaSim surrogatization architecture with a set of pre-trained components, we developed the JuliaSim Model Library and training infrastructure for large-scale surrogatization of DAE models. JuliaSim's automated model training pipeline can serve and store surrogates in the cloud. It consists of models from the Modelica Standard Library, CellML Physiome model repository \cite{yu2011physiome}, and other benchmark problems defined using ModelingToolkit.

Each of the models in the library contains a source form which is checked by continuous integration scripts, and surrogates are regenerated using cloud resources whenever the source model is updated\footnote{https://buildkite.com/}. For some models, custom importers are also run in advance of the surrogate generation. For instance, the CellMLToolkit.jl importer translates the XML-based CellML schema into ModelingToolkit.jl. Components and surrogates from other sources, such as Systems Biology Markup Language libraries (SBML), are scheduled to be generated. Additionally, for each model, a diagnostic report is generated detailing:
\begin{enumerate}
    \item the accuracy of the surrogate across all outputs of interest
    \item the parameter space which was trained on
    \item and performance of the surrogate against the original model
\end{enumerate}
is created to be served along with the models. With this information, a modeler can check whether the surrogatized form matches the operating requirements of their simulation and replace the usage of the original component with the surrogate as necessary. Note that a GUI exists for users of JuliaSim to surrogatize their own components through this same system.

% Engineers spend valuable time training locally so providing tools to avoid local training dramatically improves workflow efficiency. 

% The CellML Physiome Model Repository (PMR) \cite{yu2011physiome} is an initial test case for automatically building a library of trained surrogates. 

% This library consists of roughly 900 ODE and DAE models for physiological systems modeling.

% These models are written in the `.cellml' XML-based schema/file format. 

% Each time a model is changed and committed, our pipeline checks the git diff and then rebuilds the updated models.

% The pipeline results in a disk-saved copy of the surrogate weights for future evaluation and a diagnostics report PDF detailing the accuracy and speed improvements for the surrogate configuration, compared with the full model. 

%  ^ I'd like to include a picture of the report or something, but I get QT and weave issues that make it look quite bad. It'd be great to get this fixed (i don't think it'd be too difficult)

% \subsection{Training and Deployment Pipeline} 

% \todo[inline]{(Avik) I don't have much idea how the deployment pipeline works; might be worthwhile to get Anand and Elliot's inputs}

% \todo[inline]{Training Pipeline}

% \todo[inline]{Deployment} - Not going in this paper most likely

% JuliaSim is a software package that automates the deployment of surrogates for design, optimization and co-simulation. It supports distributed simulation of FMUs in parallel  

% \subsection{Training and Deployment Pipeline}

\section{Accelerating Building Simulation with Composable Surrogates}

To demonstrate the utility of the JuliaSim architecture, we focus on accelerating the simulation of energy efficiency of buildings. Sustainable building simulation and design involves evaluating multiple options, such as building envelope construction, Heating Ventilation, Air Conditioning and Refrigeration (HVAC/R) systems, power systems and control strategies. Each choice is modeled independently by specialists drawing upon many years of development, using different tools, each with their own strengths \cite{wetter2011view}. For instance, the equation-oriented Modelica language \cite{elmqvist1999modelica, fritzson1998modelica} allows modelers to express detailed multi-physics descriptions of thermo-fluid systems \cite{laughman2014comparison}. Other tools, such as EnergyPlus, DOE-2, ESP-r, TRNSYS have all been compared in the literature \cite{sousa2012energy, wetter2013new}. 

These models are often coupled and run concurrently to make use of results generated by other models at runtime \cite{nicolai2017co}. For example, a building energy simulation model computing room air temperatures may require heating loads from an HVAC supply system, with the latter coming from a simulation model external to the building simulation tool. Thus, integration of these models into a common interface to make use of their different features, while challenging \cite{wetter2013new}, is an important task.

While the above challenge has been addressed by FMI, the resulting coupled simulation using FMUs is computationally expensive due to the underlying numerical stiffness \cite{robertson1975some} widely prevalent in many engineering models. These simulations require adaptive implicit integrators to step forward in time \cite{wanner1996solving}. For example, building heat transfer dynamics has time constants in hours, whereas feedback controllers have time constants in seconds. Thus, surrogate models are often used in building simulation \cite{westermann2019surrogate}. 

In the following sections, we describe surrogate generation of a complex Room Air Conditioner (RAC) model, which has been exported as an FMU. We then use the surrogate to find the optimal set of design parameters over which system performance is maximized, yielding two orders of magnitude speedup over using the full model. Finally, we discuss the deployment of the surrogate in a co-simulation loop coupled with another FMU. 

% \subsection{Building Simulations}

%For  example,  some  tools  are  better  suited    for    building    envelope    simulation    (e.g.    EnergyPlus,    ESP-r), some for HVAC system simulation (e.g. Modelica, TRNSYS) and others for refrigeration systems (e.g.   DOE-2.2   refrigeration   version  49a). 

% Pure integration of several existing models into a single simulation tool may be a challenging task \cite{wetter2013new}, in particular since many tools include their own numerical solver engines and calculation algorithms. Moreover, different simulation tools manage their internal data differently, which further complicates their integration. Also, given different authors, domain experts, copyright limitations, etc.an integration into a single tool may not be meaningful or possible at all. 

\subsection{Surrogates of Coupled RAC Models}

\begin{figure*}
    \centering
    \includegraphics[width=0.8\textwidth]{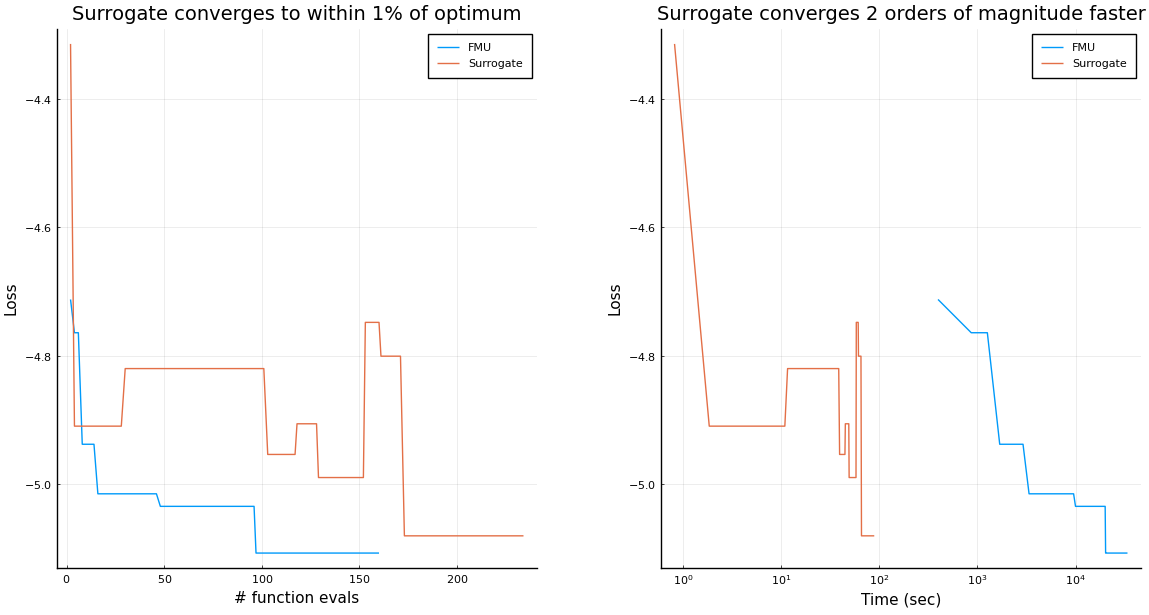}
    \caption{Comparison of global optimization while using the full model and the surrogate. Loss is measured using the full model's objective function. (Left) Convergence of loss with number of function evaluations (Right) Convergence of loss with wall clock time. The optimization using the surrogate converged much before the result from the first function evaluation of the full model is over. This is why the blue line appears translated horizontally in time.}
    \label{fig:racroom_opt}
\end{figure*}

We first consider surrogate generation of a Room Air Conditioner (RAC) model using JuliaSim, consisting of a coupled room model with a vapor compression cycle model, which removes heat from the room and dissipates it outside. The vapor compression cycle itself consists of detailed physics-based component models of a compressor, an expansive valve and a finite volume, and a staggered-grid dynamic heat exchanger model \cite{laughman2014comparison}. This equipment is run open-loop in this model to simplify the interactions between the equipment and the thermal zone. The room model is designed using components from the Modelica Buildings library \cite{wetter2014modelica}. The room is modeled as a volume of air with internal convective heat gain and heat conduction outside. The Chicago O'Hare TMY3 weather dataset\footnote{https://bcl.nrel.gov/node/58958} is imported and is used to define the ambient temperature of the air outside. This coupled model is written and exported from Dymola 2020x as a co-simulation FMU.

The model is simulated with 100 sets of parameters sampled from a chosen parameter space using Latin hypercube sampling. The simulation timespan was a full day with a fixed step size of 5 seconds. The JuliaSim FMU simulation backend runs simulations for each parameter set in parallel, and fits cubic splines to the resulting time series outputs. Then the CTESN algorithm computes projections from the reservoir time series to output time series at each parameter set. Finally, a radial basis function creates a nonlinear map between the chosen parameter space and the space of projections. Figure \ref{fig:racroom_validate} and Table \ref{tab:testtab} show the relative errors when the surrogate is tested at a parameter set on which it has not been trained. To demonstrate the reliability of the surrogate through the chosen parameter space, 100 further test parameters were sampled from the space, and the errors for each test were compiled into a histogram, as shown in \ref{fig:racroom_tests}. At any test point, the surrogate takes about 6.1 seconds to run, while the full model takes 35 minutes, resulting in a speedup of 344x. 

This surrogate model can then be reliably deployed for design and optimization, which is outlined in the following section.

% \autoref{tab:extab} illustrates the use of tables.
% It uses the \texttt{booktabs} package which provides improved typesetting of tables and \texttt{numprint} for the thousands separator.

\subsection{Accelerating Global Optimization}

Building design optimization \cite{nguyen2014review, machairas2014algorithms} has benefited from the use of surrogates by accelerating optimization through faster function evaluations and smoothing objective functions with discontinuities \cite{westermann2019surrogate, wetter2004comparison}. 

\begin{figure*}
    \centering
    \includegraphics[width = 0.8\textwidth]{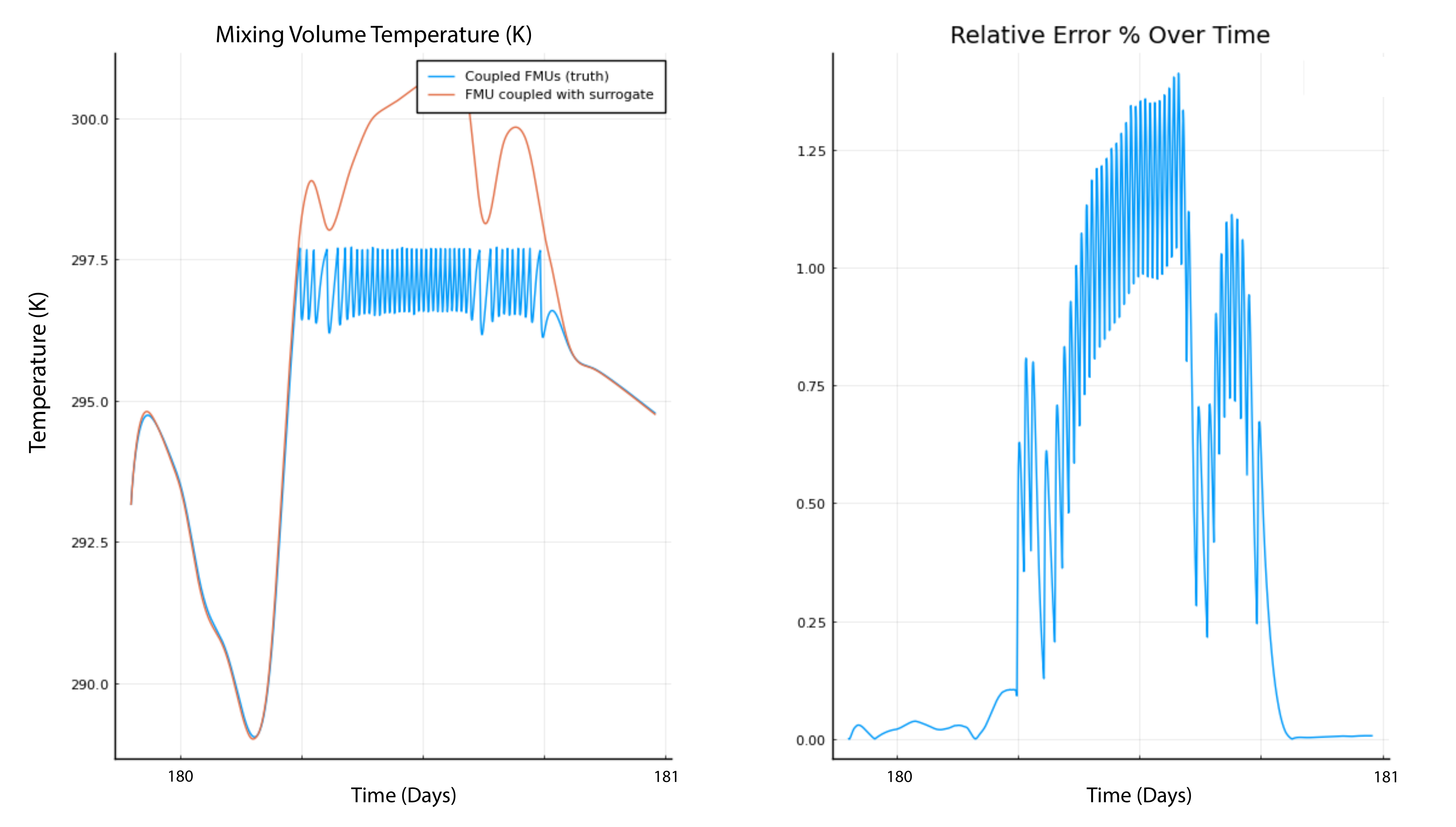}
    \caption{Coupled co-simulation of a surrogate and an FMU. The blue line represents the ground truth, which is the output from the co-simulation of two coupled FMUs, and the red line represents the output from the coupled surrogate and an FMU. While the prediction smooths over transients found in the ground truth, it does so at a relative error of less than 1.5\%. This result also empirically suggests that the output from the surrogate is bounded over the set of inputs it has received over co-simulation. The surrogate was trained over a sample of 100 inputs received from the room model. The error over the transients can be reduced by sampling more inputs from the co-simulation.}
    \label{fig:cosim}
\end{figure*}

The quantity to be maximized (or whose negative value is to be minimized) is the average coefficient of performance (COP) across the time span. We calculate this using output time series from the model by means of the following formula: 

\begin{align}
    COP(t) &= \frac{Q_{tot}(t)}{\max(0.01, CSP(t))} \\
    COP_{avg} &= \frac{\sum_{n=1}^{N_t} COP(t_n)}{N_t} 
\end{align}
where $COP$ refers to the coefficient of performance, $COP_{avg}$ refers to the average coefficient of performance across the time interval (the quantity to optimize), $Q_{tot}$ the total heat dissipation from the coupled model, $CSP(t)$ is the compressor shaft power, and $N_t$ represents the number of points in time sampled from the interval (720). 

We use an adaptive differential evolution global optimization algorithm, which does not require the calculation of gradients or Hessians \cite{price2006differential}. We chose this algorithm because of its ability to handle black-box objective functions. We use the differential optimizers in BlackBoxOptim.jl\footnote{https://github.com/robertfeldt/BlackBoxOptim.jl} for this experiment.  

Figure \ref{fig:racroom_opt} shows that the surrogate produces a series of minimizers, which eventually converge to within 1\% of the reference minimum value chosen, but two orders of magnitude faster. The surrogate does take more function evaluations to converge than the true model, but since each function value is relatively inexpensive, the impact on wall clock time is negligible. 

\subsection{Co-simulation with Surrogates}

% Surrogate models can be used to speed up cosimulation of plant Need to show an example either designing controls or a some large cosimulation

Next we examine a co-simulation loop with two coupled FMUs and replace one of the FMUs with a surrogate. Co-simulation is a form of coupled simulation where a master algorithm simulates and synchronizes time dependent models models at discrete time steps. An advantage of co-simulation over model exchange is that the individual FMUs can be shipped with their own solvers. These FMU solver calls are abstracted away from the master algorithm, which only pays heed to initialization and synchronization of the FMUs. 

We examine a simplified example of an HVAC system providing cooling to a room from the Modelica Buildings library \cite{wetter2015modelica}. Both the HVAC system and room models have been exported as FMUs, which are then imported into JuliaSim and then coupled via co-simulation. At each step of the co-simulation, the models are simulated for a fixed time step, and the values of the coupling variables are queried and then set as inputs to each other, before the models are simulated at the next time step.    

JuliaSim then generates a surrogate of the HVAC system by training over the set of inputs received during the co-simulation loop. It is then deployed in a ``plug and play'' fashion, by coupling the outputs of the surrogates to the inputs of the room and vice versa. The resultant output from the coupled system is shown in Figure \ref{fig:cosim}. The above co-simulation test has been conducted at the same set of set of design parameters as the original simulation.  \footnote{We hope to demonstrate in the final version of the paper that this surrogate can be used to explore the design space of a coupled system, by including surrogate is validated at a separate (or new) set of design points. We expect this to work but could not complete it for this initial submission.} While the individual models in this test are simplified, they serve as a proof of concept for a larger coupled simulation, either involving more FMUs or involving larger models, which may be prohibitively expensive \cite{wetter2015design}.

\section{Conclusion}

We demonstrate the capabilities of JuliaSim, a software for automated generation of deployment of surrogates for design, optimization and coupled simulation. Our surrogates can faithfully reproduce outputs from detailed multi-physics systems and can be used as stand-ins for global optimization and coupled simulation. 

While this work demonstrates an architecture capable of direct incorporating of machine learning techniques into equation-based modeling and simulation, there are many avenues for this work to continue. Further work to deploy these embedded surrogates as FMUs themselves is underway. This would allow JuliaSim to ship accelerated FMUs to other platforms. Other surrogate algorithms, such as proper orthogonal decomposition~\cite{chatterjee2000introduction}, neural ordinary differential equations~\cite{chen2018neural, kim2021stiff}, and dynamic mode decomposition~\cite{schmid2010dynamic} will be added in upcoming releases and rigorously tested on the full model library. Incorporating machine learning in other fashions, such as within symbolic simplification algorithms, is similarly being explored. But together, JuliaSim demonstrates that future modeling and simulation software does not need to, and should not, eschew all of the knowledge of the past equation-based systems in order to bring machine learning into the system.

% \begin{figure}[b]
% \centering
% \includegraphics[width=0.4 \textwidth]{resources/figure1}
% \caption{An example of a figure that fits into one column.}
% \label{fig:figure1}
% \end{figure}

% \begin{figure*}[t]
% \centering
% \includegraphics[width=0.9 \textwidth]{resources/figure2}
% \caption{Another example of a figure that spans over two columns.}
% \label{fig:figure2}
% \end{figure*}

% \autoref{tab:extab} illustrates the use of tables.
% It uses the \texttt{booktabs} package which provides improved typesetting of tables and \texttt{numprint} for the thousands separator.
% \begin{table}[htbp]
%   \caption{Sizes of compiler phases, lines of code.}\label{tab:extab}
%   \centering
%  \begin{tabular}{p{6cm}r} \toprule
%       \emph{Compiler Phase} & \emph{Lines} \\
%       \midrule
%      FrontEnd & \numprint{92192} \\
%       BackEnd & \numprint{29190} \\
%      Code generation & \numprint{8957} \\
%       \emph{Total size} & \emph{\numprint{130339}} \\
%       \bottomrule
%   \end{tabular}
% \end{table}

\section*{Acknowledgements}
The information, data, or work presented herein was funded in part by ARPA-E under award numbers DE-AR0001222 and DE-AR0001211, and NSF award number IIP-1938400. The  views  and  opinions of authors expressed herein do not necessarily state or reflect those of the United States Government or any agency thereof. 

\printbibliography

\end{document}